\newcommand{\be}{\begin{equation}}
\newcommand{\ee}{\end{equation}}
\newcommand{\bea}{\begin{eqnarray}}
\newcommand{\eea}{\end{eqnarray}}

\newcommand{\yyy}{}
\newcommand{\Pf}{{\rm Pf}}
\newcommand{\cW}{{\cal W}}
\newcommand{\bone}{{\bar{1}}}
\newcommand{\btwo}{{\bar{2}}}
\newcommand{\ba}{{\bar{a}}}
\newcommand{\bb}{{\bar{b}}}
\newcommand{\bc}{{\bar{c}}}
\newcommand{\bd}{{\bar{d}}}
\newcommand{\xx}{x}
\newcommand{\xy}{y}
\newcommand{\xw}{w}
\newcommand{\xz}{z}
\newcommand{\xp}{p}
\newcommand{\xq}{q}
\newcommand{\xX}{X}
\newcommand{\xY}{Y}
\newcommand{\xW}{W}
\newcommand{\xZ}{Z}
\newcommand{\xP}{P}
\newcommand{\xQ}{Q}
\newcommand\Ia{ {I} }
\newcommand\Jb{ {J} }
\newcommand\Kg{ {K} }
\newcommand\Ld{ {L} }
\newcommand{\voo}{v_{1\bone}}
\newcommand{\vto}{v_{2\bone}}
\newcommand{\vot}{v_{1\btwo}}
\newcommand{\vtt}{v_{2\btwo}}
\newcommand{\tQ}{{\tilde{Q}}}
\newcommand{\metric}{g}
\newcommand{\MIJ}{M_{\Ia}{}^{\Jb}}
\newcommand{\lam}{}
\newcommand{\mutil}{\tilde\mu}
\newcommand{\1}{1\kern -3pt \mathrm{l}}
\newcommand{\al}{\alpha}

\newcommand{\ep}{\epsilon}
\newcommand{\si}{\sigma}

\newcommand{\Om}{\Omega}

\newcommand{\Sp}{\mathrm{Sp}}
\newcommand{\U}{\mathrm{U}}
\newcommand{\SU}{\mathrm{SU}}
\newcommand{\SO}{\mathrm{SO}}

\newcommand{\R}{\mathrm{I}\kern -2.5pt \mathrm{R}}
\newcommand{\Z}{\mathsf{Z}\kern -5pt \mathsf{Z}}
\newcommand{\C}{\mathsf{I}\kern -5pt \mathrm{C}}

\newcommand{\Tr}{{\rm Tr}}

\newcommand{\rar}{\rightarrow}

\newcommand{\non}{\nonumber}

\newcommand{\cF}{\mathcal{F}}
\newcommand{\cN}{\mathcal{N}}

\newcommand{\half}{\mbox{$\frac{1}{2}$}}

\documentclass[12pt]{article}

\newdimen\tableauside\tableauside=1.0ex
\newdimen\tableaurule\tableaurule=0.4pt
\newdimen\tableaustep
\def\phantomhrule#1{\hbox{\vbox to0pt{\hrule height\tableaurule
width#1\vss}}}
\def\phantomvrule#1{\vbox{\hbox to0pt{\vrule width\tableaurule
height#1\hss}}}
\def\sqr{\vbox{%
  \phantomhrule\tableaustep

\hbox{\phantomvrule\tableaustep\kern\tableaustep\phantomvrule\tableaustep}%
  \hbox{\vbox{\phantomhrule\tableauside}\kern-\tableaurule}}}
\def\squares#1{\hbox{\count0=#1\noindent\loop\sqr
  \advance\count0 by-1 \ifnum\count0>0\repeat}}
\def\tableau#1{\vcenter{\offinterlineskip
  \tableaustep=\tableauside\advance\tableaustep by-\tableaurule
  \kern\normallineskip\hbox
    {\kern\normallineskip\vbox
      {\gettableau#1 0 }%
     \kern\normallineskip\kern\tableaurule}%
  \kern\normallineskip\kern\tableaurule}}
\def\gettableau#1 {\ifnum#1=0\let\next=\null\else
  \squares{#1}\let\next=\gettableau\fi\next}

\newcommand{\Yfund}{\tableau{1}}

\begin{document}

\begin{flushright}
 BRX-TH-501 \\ BOW-PH-124  \\ {\tt hep-th/0204023}
\end{flushright}
\vspace{1mm}
\begin{center}{\bf\Large\sf A cascading {\large  $\cN=1$}
{\large $\Sp(2N{+}2M){\times}\Sp(2N)$ } gauge theory }
\end{center}
\vskip 3mm
\begin{center}
Stephen G. Naculich\footnote{
Research supported in part by the NSF under grant no.~PHY94-07194 \\
\phantom{aaa}  through the ITP Scholars Program.}$^{,a}$,
Howard J. Schnitzer
\footnote{Research supported in part by the DOE under grant
DE--FG02--92ER40706.}$^{,b}$,
and Niclas Wyllard\footnote{
Research supported by the DOE under grant DE--FG02--92ER40706.\\
{\tt \phantom{aaa} naculich@bowdoin.edu;
schnitzer,wyllard@brandeis.edu}\\}$^{,b}$

\vspace*{0.3in}

$^{a}${\em Department of Physics\\
Bowdoin College, Brunswick, ME 04011}

\vspace{.2in}

$^{b}${\em Martin Fisher School of Physics\\
Brandeis University, Waltham, MA 02454}

\end{center}

\vskip 5mm

\begin{abstract}
We study the $\cN=1$ $\Sp(2N{+}2M){\times}\Sp(2N)$ cascading gauge theory on
a stack of $N$ physical and $M$ fractional (half) D3-branes
at the singularity of an orientifolded conifold. In addition to the
D3-branes and an O7-plane, the background contains eight D7-branes,
which give rise to matter in the fundamental representation of the gauge
group. The moduli space of the gauge theory is analyzed and its structure
is related to the brane configurations in  
the dual type IIB theory and in type IIA/M-theory.

\end{abstract}

\setcounter{equation}{0}
\section{Introduction} \label{Intro}

The desire to extend the original AdS/CFT correspondence
\cite{Maldacena:1998}
to examples
with less supersymmetry  has prompted the study of branes at conical
singularities. An important example is that of $N$ D3-branes at the
singularity of the conifold \cite{Klebanov:1998}.
The resulting four-dimensional
$\cN=1$ gauge theory has gauge group $\SU(N){\times}\SU(N)$ and
chiral matter multiplets in the bifundamental representations of
the gauge group. The addition of $M$ fractional D3-branes changes
the gauge group to $\SU(N{+}M){\times}\SU(N)$
\cite{Klebanov:2000a}
(other models within the same universality
class have also recently attracted attention, see
e.g.~\cite{Maldacena:2000}).
This non-conformal theory exhibits a duality cascade
\cite{Klebanov:2000b}
\be
\SU(N{+}M){\times} \SU(N) \rar \SU(N{-}M){\times} \SU(N) \rar
\ldots \rar \SU(M{+}p){\times} \SU(p) \,,
\ee
with $1\leq p\leq M$, where the simplest case is $p=1$ for which
one finds an $\SU(M{+}1){\times} \SU(1)\cong \SU(M{+}1)$ theory
at the end of the cascade.

A richer example is that of D3-branes at the singularity of an
orientifolded conifold (where the orientifold arises from an
O7-plane together with 8 D7-branes required for consistency)
leading to an $\cN=1$ $\Sp(2N){\times}\Sp(2N)$ gauge theory
with matter in the bifundamental and fundamental representations
of the gauge group \cite{Naculich:2001}.
The addition of $M$ fractional D3-branes changes the gauge group
to $\Sp(2N{+}2M){\times} \Sp(2N)$ and leads to a cascade
\be
\Sp(2N{+}2M){\times} \Sp(2N) \rar \Sp(2N{-}2M){\times} \Sp(2N)
\rar  \ldots \rar \Sp(2M{+}2p){\times} \Sp(2p) \,.
\ee
At the end of the cascade one arrives at
an $\Sp(2M+2p){\times}\Sp(2p)$ gauge theory, where $2p \le 2M$,
the simplest case being $2p=2$.

In this paper we study the $\Sp(2N{+}2M){\times} \Sp(2N)$ gauge
theory  on a stack of $N$ physical D3-branes and $M$ fractional
(half) D3-branes placed at the singularity of the orientifolded
conifold mentioned above.
This field theory is dual to type IIB string theory on
$\mathit{AdS}_5{\times}T^{11}/\Z_2$ where the $\Z_2$ is
an orientifold operation described in more detail later.
In the dual theory the $N$ D3-branes are replaced by an
$\cF_5$ flux on $T^{11}/\Z_2$ and the $M$ fractional branes
are replaced by an $F_3$ flux on an $S^3/\Z_2$ inside
$T^{11}/\Z_2$. This model, which is a natural extension
of previously studied models
\cite{Klebanov:2000b,Ahn:2001,Imai:2001},
is interesting because the D7-branes give rise to matter fields
in the field theory transforming in the fundamental representation
of the gauge group, which leads to an intricate Higgs branch
structure of the moduli space of the theory.

This paper is organized as follows. In section \ref{Ori} we
briefly review the relevant orientifolded conifold theories,
while in section \ref{Casc} we describe some aspects of the
cascade of the $\Sp(2N{+}2M){\times}\Sp(2N)$ theory and check
that the Klebanov-Strassler supergravity solution \cite{Klebanov:2000b}
is also a solution in the orientifolded theory.
Section \ref{Mod} is devoted to a study of the (classical) moduli
space of the $\Sp(2N_1){\times}\Sp(2N_2)$ gauge theory with
chiral multiplets in both the fundamental and bifundamental
representations of the gauge group.
The analysis of this section sets the stage for the more
detailed analysis in section \ref{End} of the full quantum
moduli space of the $\Sp(2M{+}2){\times}\Sp(2)$ theory at
the end of the duality cascade.
We carry out the analysis in section \ref{End} in two steps,
first describing the classical moduli space and then the quantum
moduli space.
We also discuss the interpretation of the moduli space
in terms of the dual string theory. We find that the classical and
quantum solutions join smoothly and that a deformation of the
conifold arises in the quantum theory as expected.  In
section \ref{IIA} we discuss the interpretation of the moduli
space in terms of type IIA and M-theory brane configurations.
In section 7 we summarize our findings.

\setcounter{equation}{0}
\section{Orientifolded conifold theories} \label{Ori}

The $\cN=1$ $\SU(N){\times} \SU(N)$ superconformal gauge theory
with chiral multiplets in the
$2 (\Yfund,\overline{\Yfund}) \oplus 2(\overline{\Yfund},\Yfund)$
bifundamental representations
arises as  the low energy limit of the world-volume theory
on $N$ D3-branes at a conifold singularity
\cite{Klebanov:1998,Morrison:1998}.
The conifold \cite{Candelas:1990} can be described as the subspace
of $\C^4$ defined by the equation $z_1^2 + z_2^2 + z_3^2 +z_4^2=0$.
Via a linear change of basis the conifold can also be written $xy = w z$.
The base of the conifold,
obtained by intersecting the above space with
$|z_1|^2 + |z_2|^2 + |z_3|^2 + |z_4|^2=1$, is
$T^{11}=\left [\SU(2){\times} \SU(2)\right]/ \U(1)$.
A striking example of the AdS/CFT correspondence
\cite{Maldacena:1998}
is the duality between this field theory,
and type IIB string theory on $\mathit{AdS}_5{\times} T^{11}$
\cite{Klebanov:1998}.

Orientifolds of the conifold
lead to further examples of the AdS/CFT correspondence.
The following two models
arise as the low-energy theories on the D3-branes
in a conifold background with an orientifold $\Z_2$
symmetry that does not break any supersymmetry,
and are dual to type IIB string theory
on $\mathit{AdS}_5{\times}T^{11}/\Z_2$
\cite{Naculich:2001,Imai:2001}:
\be
\label{models}
\begin{array}{lllll}
&\mathrm{(i)}& \Sp(2N){\times} \Sp(2N)\,,
\qquad &\mathrm{with} \qquad
&2(\Yfund,\Yfund)\oplus 4(\Yfund,1) \oplus 4(1,\Yfund) \,, \non \\
&\mathrm{(ii)}&
\Sp(2N) {\times} \SO(2N{+}2)\,,
\qquad & \mathrm{with}\qquad
&2(\Yfund,\Yfund) \,.
\end{array}
\ee

The form of the $\Z_2$ action on the conifold can be determined from the
corresponding IIA brane configurations (see section \ref{IIA} for
further details).

For model (i),  the action of the orientifold
on the conifold becomes \cite{Naculich:2001}
$z\leftrightarrow w$,
with $x$, $y$ invariant,
or, equivalently,
$(z_1,z_2,z_3,z_4) \rar (z_1,z_2,z_3,-z_4)$.
The fixed point set of this action
is the $w=z$ subspace of the conifold,
whose three-dimensional intersection with
$|z_1|^2 + |z_2|^2 + |z_3|^2 + |z_4|^2=1$
was called $X_3$ in ref.~\cite{Naculich:2001}.
The model thus contains an O7-plane,
and also for consistency 8 D7-branes.
The world volume of the O7-plane and D7-branes is
$\mathit{AdS}_5{\times} X_3$.

For model (ii), the orientifold action
can be shown to be
$x\rar -x$, $y\rar -y$ and $z \leftrightarrow w$,
or, equivalently, $(z_1,z_2,z_3,z_4) \rar (-z_1,-z_2,z_3,-z_4)$
using the approach of \cite{Naculich:2001}.
This result was recently obtained in \cite{Imai:2001}
using a slightly different approach.
This action has no fixed points inside $T^{11}$,
so model (ii) has no orientifold-planes or D7-branes.

For other discussions of various orientifolds of the
conifold, see e.g. \cite{Uranga:1998, Dasgupta:1998,Park:1999,Sinha:2000}.

\setcounter{equation}{0}
\section{Cascading theories} \label{Casc}

Generalizations of the orientifolded models
considered in the previous section
may be obtained by
including fractional D3-branes at the conifold singularity,
breaking the superconformal invariance.
The addition of the fractional branes increases the rank of
first factor in the product gauge groups. The resulting field theories have
gauge groups $\Sp(2N{+}2M){\times}\Sp(2N)$
and $\Sp(2N{+}2M){\times}\SO(2N+2)$, respectively.

The lack of conformal invariance causes the gauge couplings to run.
The first factor of the $\Sp(2N{+}2M){\times} \Sp(2N)$ theory
has effectively $2N_f = 4N + 4$ fields in the fundamental representation:
four from the fundamentals,
and $4N$ from the bifundamentals.
The beta function is therefore negative,
and the coupling becomes strong in the infrared.
Seiberg duality \cite{Seiberg:1995} can be used to transform this to another
weakly-coupled gauge theory. Seiberg duality relates a strongly-coupled
$ \Sp(2N_c)$
theory with $2N_f$ chiral superfields in the fundamental representation
to a weakly-coupled
$\Sp(2N_\mathrm{f} - 2N_c-4) $
theory with the same number of fundamental superfields
\cite{Intriligator:1995}.
In our case, Seiberg duality implies
\bea
\Sp(2N{+}2M){\times} \Sp(2N) \rar \Sp(2N{-}2M){\times} \Sp(2N)\,.
\eea
In the new theory, the gauge coupling of the second group factor
now becomes strong in the infrared,
leading to a second duality transformation.
This process continues, leading to a duality cascade:
\be
\Sp(2N{+}2M){\times} \Sp(2N) \rar \Sp(2N{-}2M){\times} \Sp(2N)
\rar \Sp(2N{-}2M){\times} \Sp(2N{-}4M) \rar \ldots
\ee
just as in the case of the $\SU(N{+}M) {\times} \SU(N)$ theory
\cite{Klebanov:2000b}. At the end of the cascade one arrives at
a $\Sp(2M+2p){\times}\Sp(2p)$ theory, where $2p \le 2M$.
A similar cascading phenomenon was shown for the
$\Sp(2N{+}2M){\times} \SO(2N{+}2)$ case in
refs.~\cite{Ahn:2001,Imai:2001}.

The dual supergravity solution describing the cascade of
the $\SU(N{+}M){\times}\SU(N)$ model was found in
\cite{Klebanov:2000b} (following earlier work in \cite{Klebanov:2000a});
see also \cite{Herzog:2001}.
At the end of the cascade
the conifold is replaced by its deformed version.
The solution in \cite{Klebanov:2000b} is
also a solution of the orientifolded
theory dual to the $\Sp{\times}\Sp$ gauge theory 
(for the theory dual to the $\Sp{\times}\SO$ gauge theory,
this was shown in ref.~\cite{Imai:2001}).
This follows because $F_3$ and $H_3$ change sign
under the interchange of $z$ and $w$,
whereas the metric and $F_5$ are invariant.
Combining this
with the action of $\Om(-1)^{F_L}$,
this shows that all fields are invariant 
under the orientifold projection.

Based on the properties of the supergravity solution one expects
to find the deformed conifold at the end of the flow.
To understand the geometry at the end of the flow, we
can probe the background
with a single  D3-brane as in ref. \cite{Klebanov:2000b}, i.e.~we will
assume that $p=1$ (note that the probe brane has a mirror).
To probe the theory, therefore,
we must analyze the moduli space
of the $\Sp(2M{+}2){\times}\Sp(2)$ gauge theory. This analysis
will be carried out in section \ref{End}.
First, however, we consider the more general
case of the $\Sp(2N_1){\times}\Sp(2N_2)$ gauge theory moduli space.

\setcounter{equation}{0}
\section{The $\Sp(2N_1){\times}\Sp(2N_2)$ gauge theory
moduli space} \label{Mod}

In this section, we analyze the (classical) moduli space
of the $\cN=1$  $\Sp(2N_1) {\times} \Sp(2N_2)$ gauge theory
with four chiral matter multiplets in the fundamental representation
of each factor of the gauge group,
and two in the bifundamental representation.
To obtain the $\cN=1$ superpotential for this theory,
we start with the $\cN=2$ version of the theory, turn on
(opposite sign) masses for the adjoint chiral superfields,
which breaks the supersymmetry to $\cN=1$, and integrate out
the massive fields.
This procedure is analogous to the way one obtains the
$\cN=1$ $\SU(N_1){\times} \SU(N_2)$ theory from
its $\cN=2$  cousin \cite{Klebanov:1998}.

The $\cN=2$ $\Sp(2N_1) {\times} \Sp(2N_2)$ theory can be obtained
by orientifolding the $\cN=2$  $\SU(2N_1){\times} \SU(2N_2)$  theory.
However, for both calculational and notational purposes it is convenient
to view the the  superpotential for the $\cN=2$ $\Sp(2N_1) {\times} \Sp(2N_2)$
theory as arising from that of another $\cN=2$  $\SU(2N_1){\times} \SU(2N_2)$
gauge theory, with matter hypermultiplets in both the bifundamental {\em and}
the fundamental representations, by imposing a projection on all the fields.

We therefore consider the $\SU(2N_1) {\times} \SU(2N_2)$ theory
with two $\cN=2$ vector multiplets in
the adjoint representations of $\SU(2N_1)$ and $\SU(2N_2)$ respectively, two
$\cN=2$ hypermultiplets in the bifundamental representations, and also an
additional four $\cN=2$ hypermultiplets in the fundamental
representation of each gauge group. Our notation is such that a
lower/upper index $a= 1, \ldots, 2N_1$
denotes a component in the fundamental/antifundamental
representation of $\SU(2N_1)$, and
a lower/upper index $\ba = 1, \ldots, 2N_2$
denotes a component in the fundamental/antifundamental
representation of $\SU(2N_2)$.
 In $\cN=1$ language the two $\cN=2$ vector multiplets consist of
two  vector multiplets corresponding to the two gauge groups,
and two chiral multiplets
${\phi_{1a}}^b$ and  ${\phi_{2\ba}}^\bb$.
The two $\cN=2$ bifundamental hypermultiplets
consist of two  $\cN=1$ chiral multiplets
${A_{ia}}^\bb$ $(i=1,2)$
in the $(\Yfund,\overline{\Yfund})$  of $\SU(2N_1)\times \SU(2N_2)$
and two $\cN=1$ chiral multiplets
${B_{i\ba}}^b$ $(i=1,2)$
in the $(\overline{\Yfund},\Yfund)$ representation.
In addition, the four $\cN=2$ multiplets in the fundamental representation
consist of four $\cN=1$ chiral multiplets ${Q_1^{\Ia}}_a$ $(I=1,\ldots,4)$
in the $(\Yfund,1)$
and four $\cN=1$ chiral multiplets ${\tQ_{1\Ia}^{a}}$
in the $(\overline{\Yfund},1)$,
as well as ${Q_2^{\Ia}}_\ba$ and ${\tQ_{2\Ia}^{\ba}}$
in the $(1, \Yfund)$ and $(1,\overline{\Yfund})$ respectively.

The $\cN=2$ superpotential for this theory is
\be
\label{N=2super}
\cW_{\cN=2}  = \sqrt{2}
\left\{
\Tr \left[ {\phi_{1}}      \left(  {A_{1}}  {B_{1}}  + {A_{2}}  {B_{2}} \right)
         + {\phi_{2}}      \left(  {B_{1}}  {A_{1}}  + {B_{2}}  {A_{2}} \right)
    \right]
+ \tQ_{1\Ia}   {\phi_{1}}     Q_1^{\Ia }
- \yyy \tQ_{2 \Ia}   {\phi_{2}}     Q_2^{\Ia }
\right\} \,.
\ee
We may reduce the gauge group to $\Sp(2N_1) {\times} \Sp(2N_2)$
by imposing the projections
\be
\label{spconstraint}
{\phi_{1a}}^b =  J_{ac} {J}^{bd} {\phi_{1d}}^c\,, \qquad
\qquad {\phi_{2\ba}}^\bb =  J_{\ba\bc} {J}^{\bb\bd} {\phi_{2\bd}}^\bc \,,
\ee
on the adjoint hypermultiplets (and the vector multiplets). Here $J^{ab}$ and $J^{\ba\bb}$ are the symplectic units of $\Sp(2N_1)$ and $\Sp(2N_2)$,
respectively, which are used to raise and lower indices.
Projections on the other hypermultiplet fields
\be
\begin{array}{rclcrcl}
{B_{1\ba}}^b  &=&  - J_{\ba\bc} {J}^{bd} {A_{2d}}^\bc\,, & &
{B_{2\ba}}^b   &=&    J_{\ba\bc} {J}^{bd} {A_{1d}}^\bc\,, \\[.1in]
{\tQ_{1\Jb }^{ a}} &=&
- \metric_{\Ia\Jb}
J^{ab}  {Q_1^{\Jb}}_b\,, & &
{\tQ_{2\Jb }^{\ba}} &=&
- \metric_{\Ia\Jb}
J^{\ba\bb}  {Q_2^{\Jb}}_\bb \,,
\end{array} \label{project}
\ee
result in the $\cN=2$ $\Sp(2N_1) \times \Sp(2N_2)$ gauge theory
with two $\cN=1$  chiral multiplets in the bifundamental $(\Yfund,\Yfund)$
and four $\cN=1$ chiral multiplets in each of the fundamental representations
$(\Yfund,1)$  and $(1, \Yfund)$
(as well as chiral multiplets in the adjoint
representation of the gauge group).
In subsequent calculations we use the explicit
basis choices $\metric_{IJ} = \sigma_x \otimes \1_{2\times 2}$,
and $J^{ab} = i \sigma_y \otimes \1_{N_1 \times N_1}$
(and similarly for $J^{\ba\bb}$).
In more readable matrix notation, the projections
(\ref{spconstraint}) and (\ref{project}) become
\be
\label{matrixconstraint}
\begin{array}{rclcrcl}
\phi_1 &=& J_1 \phi_1^T J_1 \,, &&
\phi_2 &=& J_2 \phi_2^T J_2 \,, \\[.1in]
B_1 &=& -J_2 A_2^T J_1\,,    &&  B_2 &=& J_2 A_1^T J_1 \,,\\ [.1in]
\tQ_{1\Ia}^T &=&  - \lam \metric_{\Ia\Jb } J_1  Q_1^{\Jb} \,, &&
\tQ_{2\Ia}^T &=&  - \lam \metric_{\Ia\Jb} J_2 Q_2^{\Jb} \,,
\end{array}
\ee
where $J^{ab} = J_1$ and $J_{ab} = J_1^{-1} = -J_1$ (and similarly for $J^{\ba\bb} = J_2$).
We could use the constraints (\ref{matrixconstraint})
to eliminate half the fields in (\ref{N=2super})
but it will be clearer to continue
to write the superpotential as (\ref{N=2super}),
with the constraints understood.

Now we include a bare mass $\mu$ for the adjoint hypermultiplets
in the superpotential
\be
\label{masses}
\cW_{\rm mass} = \mu \,  \Tr (\phi_1^2 - \phi_2^2)\,,
\ee
breaking the $\cN=2$ supersymmetry to $\cN=1$.
Taking $\mu$ to be large, we may integrate out the adjoint
fields from the superpotential, giving the quartic
superpotential
for the $\cN=1$  $\Sp(2N_1) {\times} \Sp(2N_2)$ gauge theory:
\bea
\label{N=1super}
\cW_{\cN=1}&=&-{1\over\mu} \bigg[
\Tr (A_1 B_1 A_2 B_2 - B_1 A_1 B_2 A_2) + \half \tQ_{1\Ia} Q_1^{\Jb} \tQ_{1\Jb} Q_1^{\Ia}
  - \half \tQ_{2\Ia} Q_2^{\Jb} \tQ_{2\Jb} Q_2^{\Ia}
\non\\
&&\qquad \quad +\,\tQ_{1\Ia} (A_1 B_1 + A_2 B_2) Q_1^{\Ia}
-\yyy \tQ_{2 \Ia} (B_1 A_1 + B_2 A_2) Q_2^{\Ia}
\bigg] \,.
\eea
When integrating out $\phi_1$ and $\phi_2$,
we must implement the constraint (\ref{spconstraint}),
but this will be automatic
as long as the matter hypermultiplets obey the constraints
(\ref{project}).

Since we will later be interested in the regime
where the first gauge group is strongly coupled,
we define a set of fields that are
singlets under $\Sp(2N_1)$:
\be
\label{meson}
\begin{array}{rclcrcl}
{(N_{ij})_\ba}^\bb &=& {B_{j\ba}}^c {A_{ic}}^\bb\,,  && \MIJ &=&
\tQ_{1\Ia}^a {Q_1^{\Jb}}_a  \,, \\[.1in]
 {u_{i\Ia}^\ba } &=& {\tQ_{1\Ia}^b} {A_{ib}}^\ba\,, &&
v_{i\ba}^{\Ia} &=& {B_{i\ba}}^b {Q_1^{\Ia}}_b\,,
\end{array} \qquad (i,j=1,2)
\ee
in terms of which the superpotential (\ref{N=1super}) becomes
\bea
\label{mesonsuper}
\cW_{\cN=1} &=& -{1\over\mu}  \left[
\Tr (N_{12} N_{21} - N_{11} N_{22}) +  \half \MIJ M_{\Jb}{}^{\Ia}
- \half \tQ_{2\Ia} Q_2^{\Jb} \tQ_{2\Jb} Q_2^{\Ia} \right. \non\\
&&\left. \qquad \quad + \, u_{i\Ia} v_i^{\Ia}
- \yyy \tQ_{2\Ia} (N_{11} + N_{22} ) Q_2^{\Ia}\right]\,,
\eea
where the trace is over $\Sp(2N_2)$ indices.
For later convenience we also define the fields
\be
\label{siggy}
{\si_\ba}^\bb =  {Q_2^{\Ia}}_\ba \tQ_{2\Ia}^{\bb}\,.
\ee
even though the ${Q_2^{\Ia}}_\ba$ are themselves singlets under $\Sp(2N_1)$.

The constraints (\ref{project}) imply that
the $\Sp(2N_1)$ gauge-invariant fields obey
\bea
\label{Nconstraint}
N_{11} =  J_2 N_{22}^T J_2\,, \qquad \qquad
N_{12} &=& -J_2 N_{12}^T J_2 \,, \qquad \qquad
N_{21} = -J_2 N_{21}^T J_2 \,,\\ [.1in]
\label{uvconstraint}
u_{i\Ia}^T  &=& \epsilon_{ij} \metric_{\Ia\Jb } J_2 v_j^{\Jb }\,,
\\[.1in]
\label{Mconstraint}
\MIJ
&=& - \metric_{\Ia\Kg} \metric^{\Jb\Ld} M_{\Ld}{}^{\Kg}\,,
\\[.1in]
\label{sconstraint}
\si &=& J_2 \si^T J_2\,.
\eea
where $\epsilon_{12}=1$.
The $4\times 4$ matrix $\MIJ $ parametrized by
\be
\label{Mparam}
M = \pmatrix{-\xW    & 0    & -\xY   & \xP   \cr
                0    & \xW  & -\xQ  & -\xX \cr
              \xX    &-\xP  &  \xZ   & 0   \cr
              \xQ    & \xY  &    0   &-\xZ}
\ee
automatically satisfies the constraint (\ref{Mconstraint}).

The classical F-term equations are obtained
by varying the superpotential (\ref{N=1super})
with respect to the independent variables $A_i$ and $Q_i$
(recall that $B_i$ and $\tQ_i$ are not independent variables,
cf.~(\ref{project})). However, it is easy to see that one
obtains the same equations by treating $A_i$, $B_i$, $Q_i$,
and $\tQ_i$  as independent when performing the variation.
Varying with respect to $A_1$ and $A_2$ gives
\bea
\label{FtermA}
N_{21} B_2 - N_{22} B_1 + v_1^{\Ia} \tQ_{1\Ia}
- \yyy \si B_1 &=&0\,,\non\\
N_{12} B_1 - N_{11} B_2 + v_2^{\Ia} \tQ_{1\Ia}
- \yyy \si B_2&=&0 \,.
\eea
Multiplying these equations on the right by $A_i$, we obtain
\bea
\label{Ftermplus}
N_{21} N_{12} - N_{22} N_{11}  + v_1^{\Ia} u_{1\Ia}
- \yyy \si N_{11} &=& 0 \,,\non\\
N_{21} N_{22} - N_{22} N_{21}  + v_1^{\Ia} u_{2\Ia}
- \yyy \si N_{21} &=& 0 \,,\non\\
N_{12} N_{11} - N_{11} N_{12} + v_2^{\Ia} u_{1\Ia}
- \yyy \si N_{12}  &=& 0 \,,\non\\
N_{12} N_{21} - N_{11} N_{22} + v_2^{\Ia} u_{2\Ia}
- \yyy \si N_{22} &=& 0\,.
\eea
The F-term equations obtained by varying (\ref{N=1super})
with respect to $B_1$ and $B_2$,
\bea
\label{FtermB}
A_2 N_{12} - A_1 N_{22} + Q_1^{\Ia} u_{1 \Ia}
- \yyy A_1 \si &=& 0 \,, \non\\
A_1 N_{21} - A_2 N_{11} + Q_1^{\Ia} u_{2 \Ia}
- \yyy A_2 \si &=& 0 \,,
\eea
are equivalent to (\ref{FtermA}), using the constraints (\ref{project}).
Varying the superpotential with respect to  $Q_1$ and $\tQ_1$ yields
\bea
\label{FtermQone}
u_{1 \Ia}  B_1 + u_{2 \Ia}  B_2
+ \MIJ \tQ_{1 \Jb} &=& 0 \,, \non\\
A_1 v_1^{\Ia}  + A_2 v_2^{\Ia}
+     Q_1^{\Jb} M_{\Jb}{}^{\Ia} &=& 0 \,,
\eea
where the second equation follows from the first using (\ref{project}).
Finally,
\bea
\label{FtermQtwo}
 (N_{11}  + N_{22} + \yyy \si) Q_2^{\Ia} &=& 0 \,, \non\\
 \tQ_{2 \Ia} (N_{11}  + N_{22} +\yyy \si) &=& 0 \,,
\eea
follow by varying with respect to  $Q_2$ and $\tQ_2$ (the second equation follows
{}from the first using (\ref{project})).

When both
${A_{ia}}^\ba Q_{2 \ba}^I \tQ_{2 I}^{\bb} $ and
$Q_{1 a}^I\tQ_{1 I}^{b}  {A_{ib}}^\bb$
vanish, the
F-term equations (\ref{Ftermplus}), and the corresponding equations
that follow from (\ref{FtermB}),
imply that the set of $2N_2 {\times} 2N_2$ matrices $N_{ij}$
mutually commute,
and hence they can be diagonalized.
The eigenvalues can therefore be interpreted as the positions
of the D3-branes.
By virtue of (\ref{Ftermplus}),
\be \label{Nconi}
N_{21} N_{12} - N_{22} N_{11}  =0\,,
\ee
so these D3-branes live on a conifold.

For the unorientifolded $\SU(N{+}M){\times}\SU(N)$ model
and for the $\Sp(2N{+}2M){\times}\SO(2N{+}2)$
orientifolded theory, equation (\ref{Nconi})
describes the entire classical moduli space.
However, for the $\Sp(2N_1){\times}\Sp(2N_2)$ theory the moduli space has
additional structure.
One way to ensure that
${A_{ia}}^\ba Q_{2 \ba}^I \tQ_{2 I}^{\bb} $ and
$Q_{1 a}^I\tQ_{1 I}^{b}  {A_{ib}}^\bb$
both vanish is to set $Q_1$ and $Q_2$ to zero, but there are also other
solutions. As an example, let us assume that
$v_{i I}^b = \tQ_{1 I}^{b}  {A_{ib}}^\bb$
and choose a basis such that
$Q_{2\ba}^I$ is only non-zero for the first four entries
($\ba=1,2,3,4$, say). Let us also assume that the $N_{ij}$'s
are block diagonal with one $4{\times}4$-dimensional block and
one $(2N_2{-}4){\times}(2N_2{-}4)$-dimensional block
(it is not clear whether all solutions have this block-diagonal form).
In this case the
F-term equations split into two parts. For the
$(2N_2{-}4){\times}(2N_2{-}4)$-dimensional block it follows as
above that the $N_{ij}$'s commute; hence the eigenvalues
in this sector satisfy the conifold equation (\ref{Nconi}).
For the $4{\times}4$-dimensional block it follows from
(\ref{FtermQtwo}) that if $\si=0$ then $N_{11}+N_{22}=0$ has
to hold (assuming that the $Q_2^I$'s span the $4{\times}4$ space).
As we will see in more detail in the next section, $N_{11}+N_{22}=0$
corresponds in the dual type IIB geometry to the point where
the O7-plane and the 8 D7-branes are localized.
The implications of this solution is that when the $Q_2^I$'s are
non-zero, four of the D3-branes are stuck to the D7-branes.
When $\si$ is not zero the generic solution to
eq.~(\ref{FtermQtwo}) is given by $\si = -N_{11}-N_{22}$.
Inserting this relation into (\ref{Ftermplus}) leads to the equations
\bea
N_{21}N_{12} + N_{11}^2 = 0\,, & & N_{12}N_{21} + N_{22}^2 = 0 \,, \non \\
N_{21}N_{22} + N_{11}N_{21} =0 \,, && N_{12}N_{11} + N_{22}N_{12} = 0 \,.
\eea
It can be shown that there exist $4{\times}4$-dimensional matrices
satisfying these equations which are not mutually commuting.
The interpretation
of this non-commutative solution on the string theory side is unclear.
Since  the non-zero
$Q_2^I$'s only affect a $4{\times}4$-dimensional subspace,
they are  essentially a $1/N$ effect.
Perhaps the general framework discussed in
\cite{Berenstein:2000} can be used to shed
some light on this sector of the moduli space.

So far we have only analyzed the classical moduli space. In general
there are quantum corrections to the classical moduli space and some
solutions may not have counterparts in the full quantum moduli space.
The quantum modification of the superpotential for the
$\Sp(2N_1){\times}\Sp(2N_2)$ theory is not known. However, for the theory
at the end of the cascade, it is possible,
with certain assumptions, to determine
the quantum superpotential. In the next section we will study the
full quantum moduli space for the theory at the end of the cascade.

\setcounter{equation}{0}
\section{The $\Sp(2N_1){\times}\Sp(2)$ moduli space} \label{End}

At the end of the cascade, we have an $\Sp(2N_1) {\times} \Sp(2)$ gauge theory.
For this case, the $2{\times} 2$ matrices $N_{ij}$
satisfying (\ref{Nconstraint})
can be explicitly parametrized as
\be
\label{Nparam}
N_{11} = \pmatrix{ \xw &\xp \cr \xq & \xz}\,,\quad
N_{12} = \pmatrix{ -\xx &  0 \cr   0 &  -\xx}\,,\quad
N_{21} = \pmatrix{ \xy &  0 \cr   0 &  \xy}\,,\quad
N_{22} = \pmatrix{ -\xz &\xp \cr \xq & -\xw}\,.\quad
\ee
These satisfy
\be
N_{11} N_{22} - N_{12} N_{21} =\pmatrix{
\xx\xy - \xw \xz + \xp \xq &  0 \cr
                        0 & \xx\xy - \xw \xz + \xp \xq } \,,
\ee
and mutually commute
\be
\label{commute}
[N_{ij}, N_{kl}] =0\,.
\ee
{}From this result it follows that for the theory at the end of the cascade
there are no non-commutative solutions of the type
discussed at the end of sec.~4.

We now analyze the moduli space of this theory,
first considering the classical moduli space,
then turning to the quantum modifications due
to the dynamically-generated superpotential.

\subsection{Classical moduli space}

We do not consider the most general case,
but rather analyze regions of the moduli space where,
roughly speaking, the scalar vev of one or the other
(or both) of the fundamental fields $Q_1$ and $Q_2$ vanishes.

\bigskip
\noindent{{\bf Case I}: $v_i^\Ia =0$  and $Q_2^\Ia=0$}
\medskip

First we consider solutions of the F-term equations for which
both $v^{\Ia}_i = B_i Q_1^{\Ia}=0$  and $Q^{\Ia}_2=0$
(thus $\si=0$).
The constraints (\ref{uvconstraint}) and (\ref{project})
then imply $u_{iI}=0$ and $\tQ_{2\Ia}=0$.
The F-term equations (\ref{Ftermplus}) reduce to
\be
\label{Nconifold}
N_{11} N_{22} - N_{12} N_{21} = 0 \,.
\ee
We may
use an $\Sp(2)$ gauge transformation to diagonalize
eqs.~(\ref{Nparam}), corresponding to setting $\xp = \xq=0$.
The eigenvalues of $N_{ij}$ then correspond to the position
of the D3-brane probe and its orientifold mirror.
Eq.~(\ref{Nconifold}) implies
\be
\label{conifold}
\xx \xy - \xw \xz  = 0\,,
\ee
so the probe brane (and its mirror) move on a conifold.
Moreover, the orientifold action on the conifold described
in sec.~2,
$\xz \leftrightarrow \xw$, $\xx \to \xx$, $\xy \to \xy$,
exchanges the positions of the probe and its mirror,
so our choice of parametrization (\ref{Nparam})
is consistent with the variables used for the geometry in sec.~2.

The simplest way to satisfy $B_i Q_1^{\Ia}=0$
is to set $Q_1^{\Ia}=0$, in which case
$ \MIJ $ vanishes.
$ \MIJ $ may, however, be non-zero
if not all the $Q^{\Ia}_1$ vanish.
Multiplying the first F-term equation in (\ref{FtermQone})
on the right by $Q_1^{\Kg}$,
we obtain
\be
\MIJ M_{\Jb}{}^{\Kg}  = 0\qquad  \Rightarrow \qquad \det  M  =  0\,,
\ee
which implies
\be
\label{classicalMsoln}
\xX \xY - \xW \xZ  =0\,,\qquad \xW = \xZ\,,\qquad  \xP=\xQ=0 \,,
\ee
in terms of the parametrization (\ref{Mparam}).

\bigskip
\noindent{{\bf Case II}: $v_i^{\Ia}=0$ }
\medskip

Next we consider the case where $v_i^{\Ia}=B_i Q_1^{\Ia}=0$,
but some of the $Q^{\Ia}_2$ are non-vanishing.
The constraints (\ref{Nconstraint}) and (\ref{sconstraint}) imply
$ N_{11} + N_{22} + \yyy \si = J_2 (N_{11} + N_{22} + \yyy \si)^T J_2 $.
Consequently $ N_{11} + N_{22} + \yyy \si$
is proportional to a linear combination of the Pauli matrices,
and therefore is invertible if it does not vanish.
If it is invertible, then eq.~(\ref{FtermQtwo}) implies
$Q^{\Ia}_2=0$, contrary to assumption.
Therefore, it vanishes:
\be
\label{sigNN}
\yyy \si = - N_{11}-N_{22}\,.
\ee
Setting $v_i=0$ in  eq.~(\ref{Ftermplus}),
and using eqs.~(\ref{commute}) and (\ref{sigNN}),
we see that
\bea
\label{ctwo}
N_{11} N_{22} - N_{12} N_{21} &=& 0\,,\non\\  [.1in]
N_{11} +N_{22} &=& 0\,,\non\\ [.1in]
\si  &=&0\,,
\eea
which implies
\be
\label{D3loc}
\xx\xy-\xw\xz = 0,\qquad
\xw = \xz, \qquad
\xp=\xq=0 \,,
\ee
so the probe brane moves on the $\xw = \xz$ subspace
of the conifold (\ref{conifold}).
The restriction to this subspace occurs only because some of the
$Q_2^{\Ia}$ have non-zero vevs.
By comparing with the results in section \ref{Ori} we see that
the D3-brane probe (\ref{D3loc}) is stuck to the D7-branes which are
located at the orientifold fixed point, $z=w$,
so the minimal length of D3-D7 strings vanishes.
This is consistent with the fact that
the induced masses of the $Q_2^{\Ia}$ fields,
which are given by the eigenvalues of $\si$,
are zero in this case, since $\si$ vanishes identically.

The fields $Q_1^{\Ia}$ may also have non-zero vevs,
as long as they satisfy
$B_i Q_1^{\Ia}=0$ and $ \MIJ  M_{\Jb}{}^{\Kg}=0$.
The latter condition implies that $\MIJ$ satisfies eq.~(\ref{classicalMsoln}).

\bigskip
\noindent{{\bf Case III}: $Q^\Ia_2=0$ }
\medskip

Finally, we consider the case  in which
$Q_2^{\Ia}=0$ (therefore  $\tQ_{2\Ia}=0$),
but some of the $Q_1^{\Ia}$ are nonzero.
Setting $\si=0$ in eqs.~(\ref{Ftermplus}),
and using (\ref{commute}), we see that
\be
\label{ortho}
v_{j\ba}^{\Ia}  u_{i\Ia}^\bb \propto \delta_{ij} {\delta_\ba}^\bb\,.
\ee
We assume that the constant of proportionality does not  vanish,
otherwise this reduces to case I.
Viewing $ {v_j^{\Ia}}_\ba $ as vectors whose components are labelled by
$\Ia$
we choose a basis in which
\bea
\label{vparam}
{v_{1}^{\Ia}}_\bone = \pmatrix{\voo \cr 0\cr 0 \cr 0}, \qquad
{v_{2}^{\Ia}}_\btwo = \pmatrix{0 \cr \vtt \cr 0 \cr 0}, \qquad
{v_{2}^{\Ia}}_\bone = \pmatrix{0 \cr 0 \cr \vto \cr 0}, \qquad
{v_{1}^{\Ia}}_\btwo = \pmatrix{0 \cr 0 \cr 0 \cr \vot}. \qquad
\eea
The constraints (\ref{uvconstraint}) then imply
\bea
\label{uparam}
u_{1\Ia}^{\bone} = \pmatrix{\lam \vtt \cr 0\cr 0 \cr 0}, \qquad
u_{2\Ia}^{\btwo} = \pmatrix{0 \cr \lam \voo \cr 0 \cr 0}, \qquad
u_{2\Ia}^{\bone} = \pmatrix{0 \cr 0 \cr -\lam \vot \cr 0}, \qquad
u_{1\Ia}^{\btwo} = \pmatrix{0 \cr 0 \cr 0 \cr -\lam \vto}, \qquad
\eea
and the relations (\ref{ortho}) imply
\be
\vto \vot = - \voo \vtt\,.
\ee
The F-term equations (\ref{Ftermplus}) then give
\be
\label{Ncon}
{(N_{11} N_{22} - N_{12} N_{21})_\ba}^\bb =
\lam \voo \vtt {\delta_\ba}^\bb \,.
\ee
Multiplying the first equation of (\ref{FtermQone}) on the right
by ${A_{ja}}^\bb$ and on the left by ${v_i^{\Ia}}_\ba$ we get
\be
\label{NMrelation}
\lam \voo\vtt {\left( N_{ji}\right)_\ba}^\bb
+ {v_i^{\Ia}}_\ba \MIJ u_{j\Jb}^\bb  = 0 \,.
\ee
Using (\ref{Nparam}), (\ref{vparam}), and (\ref{uparam}),
this can be used to show that
$M$ has the form (\ref{Mparam}) with
\be
\label{equality}
\xX={\voo \over \vto} \xx\,, \quad \;\;
\xY={\vto \over \voo} \xy\,, \quad \;\;
\xW=\xw\,, \quad \;\;
\xZ=\xz\,, \quad \;\;
\xP = -{\vot\over\voo} \xp\,, \quad \;\;
\xQ = -{\voo\over\vot} \xq\,.
\ee
Next,
we multiply eqs.~(\ref{FtermA}) and (\ref{FtermB}) by $Q_1$ and $\tQ_1$,
eq.~(\ref{FtermQone}) by $A_i$ and $B_i$,
and compare the results to show
\bea
\label{orbifold}
N_{11} + N_{22} &=& 0 \,.
\eea
Equation (\ref{orbifold}) arises only when
the $Q_1^\Ia$ vevs are not all zero.
Equations (\ref{Ncon}) and (\ref{orbifold}) imply
\be
\label{newD3loc}
\xx\xy - \xw \xz  =  \lam \voo \vtt,
\qquad \xw=\xz, \qquad \xp =\xq = 0 \,.
\ee
The matrix $\MIJ$ is
completely determined in terms
of the $v_i^\Ia$ and $N_{ij}$ as
\be
\label{Mclassical}
M = \pmatrix{ -\xz    & 0    & -\vto \xy/\voo   & 0   \cr
                0    &\xz  &    0   & -\voo\xx/\vto \cr
             \voo \xx/\vto    & 0   & \xz   & 0   \cr
            0 & \vto\xy/\voo   & 0   & -\xz}
\ee
and obeys $\det M = (\voo \vtt)^2$.

The geometrical interpretation of eq.~(\ref{newD3loc})
is not entirely clear.
The induced masses of the $Q_1^{\Ia}$ fields,
which are given by the eigenvalues of $M$,
are nonvanishing when $\voo \vtt \neq 0$.
This would appear to imply that the length
of the D3-D7 strings in this case is nonvanishing. 

It would be interesting to find the generalization 
of the solution in \cite{Klebanov:2000b} describing this 
sector of the moduli space.

\subsection{Quantum moduli space}

At the end of the flow, the first
gauge group of the $\Sp(2N_1){\times} \Sp(2)$
 theory becomes strongly coupled,
and a quantum superpotential is dynamically generated.
We effectively have an $\cN=1$ $\Sp(2N_1)$ gauge
theory with $2N_f = 8$ hypermultiplets $q^M_a$, which we parametrize as
\be
\label{qbasis}
q^M_a =  \pmatrix{  {A_{1a}}^\bone & {A_{1a}}^\btwo &
			{A_{2a}}^\bone & {A_{2a}}^\btwo &
			Q_{1a}^{1} & Q_{1a}^{2} &
                        Q_{1a}^{3} & Q_{1a}^{4} } \,.
\ee
The gauge indices of the $\Sp(2N_2)$ factor act as
flavor indices.
When the $\cN=2$ superpotential for such a theory
has the form $\sqrt{2} q^M_a \phi^{ab} q^M_b$,
where  $M=1,\cdots, 2N_f$,
the Affleck-Dine-Seiberg superpotential \cite{Affleck:1984} is given by
\cite{Intriligator:1995, deBoer:1998} (when $N_1{+}1>N_f$)
\be
\label{ADS}
\cW_{\rm ADS} = (N_1 + 1 - N_f)
\left(\Lambda^{3(N_1+1)-N_f}_{\cN=1} \over \Pf~ V\right)^{1\over N_1+1-N_f} \,,
\ee
where $V$ is the antisymmetric $2N_f {\times} 2N_f$ meson matrix
$V^{MN}  = q^M_a J^{ab} q^N_b$.
The $\cN=2$ superpotential (\ref{N=2super})
for the $\Sp(2N_1)\times \Sp(2)$ theory is not flavor diagonal
in the basis (\ref{qbasis}), but can be written as
$ \sqrt{2} q^M_a g_{MN} \phi^{ab} q^N_b $ where
\be \label{metric}
g_{MN} = \pmatrix{ 0 & 0 & 0 & -1 & 0 & 0 & 0 & 0 \cr
                   0 & 0 & 1 &  0 & 0 & 0 & 0 & 0 \cr
                   0 & 1 & 0 &  0 & 0 & 0 & 0 & 0 \cr
                  -1 & 0 & 0 &  0 & 0 & 0 & 0 & 0 \cr
                   0 & 0 & 0 &  0 & 0 & \lam 1 & 0 & 0 \cr
                   0 & 0 & 0 &  0 & \lam 1 & 0 & 0 & 0 \cr
                   0 & 0 & 0 &  0 & 0 & 0 & 0 & \lam 1 \cr
                   0 & 0 & 0 &  0 & 0 & 0 & \lam 1 & 0  } \,,
\ee
where the lower $4\times 4$ block is just the matrix $g_{IJ}$
introduced in eqs.~(\ref{project}).
However, $g_{MN}$ can be diagonalized by a change of basis
without altering the Pfaffian.
Hence, the superpotential of the $\cN=1$
$\Sp(2N_1)\times \Sp(2)$ theory can be written as
\be
\label{fullsuper}
\cW = \cW_{\cN=1} + (N_1 - 3)
\left(\Lambda^{3N_1-1}_{\cN=1} \over \Pf~ V\right)^{1\over N_1-3} \,,
\ee
with $\cW_{\cN=1}$ given by eq.~(\ref{mesonsuper})
and with $V$ given by
\be
\label{mesonmatrix}
V^{MN}  = q^M_a J^{ab} q^N_b
       = \pmatrix{  0 & \xx & -\xq & \xw & 0 & -\vtt & 0 & 0 \cr
                  -\xx & 0    & -\xz  & \xp & 0 &     0 & \vto & 0 \cr
                  \xq & \xz & 0    & \xy & 0 &     0 & 0 & \vot \cr
                 -\xw & -\xp & -\xy &   0 & -\voo & 0 & 0 & 0 \cr
      0    &    0 &    0 &\voo & 0      & \lam\xW &-\lam\xQ & \lam -\xX \cr
      \vtt &    0 &    0 &   0 &-\lam\xW&       0 &-\lam\xY & \lam \xP \cr
      0    &-\vto &    0 &   0 & \lam\xQ& \lam\xY &       0 & -\lam \xZ \cr
      0    &    0 &-\vot &   0 & \lam\xX&-\lam\xP & \lam\xZ &         0 } \,,
\ee
where we have chosen the basis (\ref{vparam}) and (\ref{uparam})
for ${v_{i}^{\Ia}}_\ba$ and $ u_{i\Ia}^{\ba}$,
and used the parametrizations (\ref{Nparam}) and (\ref{Mparam}).
Using the same parametrization, the superpotential (\ref{fullsuper})
becomes
\bea
\label{fullsuperexp}
\cW& =& - {1\over \mu} \left[
 2 (-\xx\xy +\xw\xz -\xp \xq) + 2 \lam (\voo \vtt  - \vot \vto)
+ \xW^2  + \xZ^2 - 2 \xX \xY + 2 \xP \xQ \right. \non\\
&&\left. \qquad +  \yyy \, (\xz - \xw) ({\si_\bone}^\bone - {\si_\btwo}^\btwo)
- \yyy 2 \xp {\si_\btwo}^\bone - \yyy 2 \xq {\si_\bone}^\btwo
 - \half \tQ_{2\Ia} Q_2^{\Jb} \tQ_{2\Jb} Q_2^{\Ia} \right] \non\\
&& \qquad + \,(N_1 - 3) \left(\Lambda^{3N_1-1}_{\cN=1} \over \Pf~ V\right)^{1\over N_1-3} \,,
\eea
with
\bea
\label{Pfexp}
&&\Pf (V) =  \sqrt{ \det V}
= ( \xx \xy - \xw \xz +\xp \xq) (\xX \xY - \xW \xZ +\xP \xQ)
-\voo \vot \vto \vtt \non\\
&&+ \lam \left( \xx \xY \voo \vot - \xy \xX \vto \vtt
        - \xw \xW \vot \vto +  \xz \xZ \voo \vtt
        + \xq \xP \voo \vto -  \xp \xQ \vot \vtt \right) \,.
\eea

The F-term equations are derived from the superpotential (\ref{fullsuperexp})
by varying with respect to the gauge invariant fields
$N_{ij}$, $\MIJ$, $v^I_i$, and  $Q_2^\Ia$.
These equations differ from the classical F-term equations
(\ref{FtermA}), (\ref{FtermB}), and (\ref{FtermQone}),
even in the limit $\Lambda_{\cN=1} \to 0 $,
because the latter were obtained by varying (\ref{N=1super})
with respect to $A_i$, $B_i$ and $Q_1^\Ia$.
The F-term equations (\ref{FtermQtwo}),
obtained by varying with respect to $Q_2^\Ia$,
are the same in the classical and quantum cases,
because the ADS superpotential does not depend on $Q_2$.
Even though their derivations are different we will find
that the quantum and classical solutions join smoothly.

\bigskip
\noindent{{\bf Case I}: $v^\Ia_i=0$  and $Q^\Ia_2=0$}
\medskip

As a simplification we can set $v^I_i=0$ and $\si=0$ directly in
(\ref{fullsuperexp}), (\ref{Pfexp})
since these expressions are quadratic in $v_i^I$
and $Q_2^I$ and hence will not
contribute to the variation.
Comparing the F-term equations derived by varying
(\ref{fullsuperexp}) with respect to
$\xx$, $\xy$, $\xw$, $\xz$, $\xp$, and $\xq$,
and with respect to
$\xX$, $\xY$, $\xW$, $\xZ$, $\xP$, and $\xQ$,
we obtain
\bea
\label{qone}
\xx\xy-\xw\xz+\xp\xq  &=&\xX\xY-\xW\xZ +\xP \xQ \,,\non\\
\xW  & = & \xZ  \,,\non\\
\xP = \xQ &= &0 \,,
\eea
implying $\Pf(V) = (\xx\xy-\xw\xz+\xp\xq)^2$.
The F-term equations become
\be
\label{qoneFterm}
-{2\over\mu} - \Lambda^{3N_1-1 \over N_1-3}_{\cN=1} (\Pf~ V)^{2-N_1 \over N_1-3}
(\xx\xy-\xw\xz-\xp\xq) = 0 \,,
\ee
which implies
\be
\label{qonedetN}
{(N_{11} N_{22} - N_{12} N_{21})_\ba}^\bb
= \epsilon \delta_\ba^\bb\,,
\qquad {\rm where} \qquad
\epsilon =
\left(\mu \over 2\right)^{N_1-3 \over N_1-1}
\Lambda^{3N_1 - 1 \over N_1 - 1}_{\cN=1} \,.
\ee
Setting $\xp=\xq=0$ using  an $\Sp(2)$ gauge transformation,
we find that the probe branes move on a deformed conifold
\be
\label{deformedconifold}
\xx\xy-\xw\xz =    \epsilon.
\ee
{}From (\ref{qone}), the matrix $\MIJ$ has the form
\be
\label{Mquantum}
M_I{}^J = \pmatrix{ -\xZ    & 0    & -\xY   &  0   \cr
                0    & \xZ  &   0   & -\xX \cr
              \xX    &   0  &  \xZ   & 0   \cr
                 0   & \xY  &    0   & -\xZ},
\ee
where the matrix elements of $M$ are arbitrary, but
by (\ref{qone}) and (\ref{deformedconifold}) must satisfy
\be
\label{Meqn}
\det M = (\xX\xY - \xZ^2)^2 = \epsilon^2 \,.
\ee
Unlike in the classical case, $M =0$ is not a solution.
(If $\MIJ$ were to vanish, then $\cW_{\rm ADS}$ would blow up.)
The lower $4\times 4$ block of the antisymmetric matrix $V$ has the form
\be \label{Vmatrix}
V^{IJ} =  Q_{1a}^I J^{ab} Q_{1b}^J = \metric^{IK} M_{K}{}^{J}
= \pmatrix{   0      & \lam\xZ &       0 &-\lam \xX \cr
                    -\lam\xZ&       0 &-\lam\xY &        0 \cr
      	                   0& \lam\xY &       0 & -\lam \xZ \cr
                     \lam\xX&       0 & \lam\xZ &         0 }\,.
\ee
A flavor transformation allows us to block-diagonalize this matrix,
so that the above relations reduce to
\be
\label{deBoersoln}
V^{IJ} = \pmatrix{   0      & \lam\tilde{Z} &       0 &        0 \cr
                    -\lam\tilde{Z}&       0 &       0 &        0 \cr
      	                   0&       0 &       0 &-\lam \tilde{Z} \cr
                           0&       0 & \lam\tilde{Z} &         0 } \,,
\qquad {\rm with} \qquad
\tilde{Z} = \epsilon^{1\over 2} =
\left(\mu \over 2\right)^{N_1-3 \over 2N_1-2}
\Lambda^{3N_1 - 1 \over 2N_1 - 2}_{\cN=1} \,,
\ee
which is exactly the meson matrix
in eq. (3.12) of de Boer et.~al.~\cite{deBoer:1998}
for the $\cN=1$ $\Sp(2N_1)$ theory with $2N_f=4$ fundamental fields
(see also ref.~\cite{Ahn:1998}).
In section 6, we will explain this
in terms of the M-theory configuration corresponding
to this branch of moduli space.
Note that (\ref{deBoersoln}) is simply a rewriting of
(\ref{Meqn}) since the determinant is invariant under the flavor rotation.

In the limit $\Lambda_{\cN=1} \to 0$,
the solution (\ref{deformedconifold}) and (\ref{qone})
reduces to
the classical solution (\ref{conifold}) and (\ref{classicalMsoln}).

\bigskip
\noindent{{\bf Case II}: $v^\Ia_i=0$}
\medskip

The F-term equation obtained by varying
the full superpotential (\ref{fullsuperexp}) with respect to $Q_2$
is equivalent to the classical F-term equation (\ref{FtermQtwo}).
By the previous arguments given for the classical case II  above,
this yields
\be
\yyy \si = - N_{11}-N_{22}\,.
\ee
The F-term equations derived by varying (\ref{fullsuperexp}-\ref{Pfexp})
with respect to  $N_{ij}$  and $\MIJ$, after setting $v^\Ia_i =0$,
yield (\ref{qone}-\ref{qoneFterm}) as in case I.
In addition, they imply
\be
{\si_\ba}^\bb \propto \delta_\ba^\bb \,.
\ee
This, together with the constraint (\ref{sconstraint}),
implies that $\si$ vanishes.
Hence we have
\bea
\label{qtwo}
N_{11} N_{22} - N_{12} N_{21} &=&   \epsilon \1 \,, \non\\[.1in]
N_{11} +N_{22} &=& 0\,,\non\\[.1in]
\si &=&0 \,.
\eea
The second equation in (\ref{qtwo}) implies $\xw=\xz$ and $\xp=\xq=0$,
so the probe brane moves on the $\xw = \xz $
subspace  of the  deformed conifold (\ref{deformedconifold}).
As in case I, the field $M$ is of the form
(\ref{Mquantum}) satisfying (\ref{Meqn}).

When $\Lambda_{\cN=1} \to 0$,
the quantum case II solution (\ref{qtwo})
reduces to the classical case II solution (\ref{ctwo}).

\bigskip
\noindent{{\bf Case III}: $Q^\Ia_2=0$ }
\medskip

Setting $\si=0$ in eq.~(\ref{fullsuperexp}),
and varying with respect to $N_{ij}$, $\MIJ$, and $v_{i\ba}$
we find
\bea
\label{qthree}
\xx \xy - \xw \xz &=&  \lam \voo\vtt  +  \epsilon \non\\
\xw  &=& \xz   \non\\
\xp = \xq &=& 0  \non\\
\vot \vto &=& -\voo \vtt.
\eea
Thus
\bea
\label{qthreemat}
N_{11} N_{22} - N_{12} N_{21}
&=&(\lam\voo\vtt +  \epsilon ) \1 \non\\[.1in]
N_{11} +N_{22} &=& 0
\eea
The F-term equations also show that the matrix elements of
$\MIJ$ are related to those of $N_{ij}$ by
\be
\label{related}
\xX = \left( \voo\over\vto \right) \xx\,, \qquad
\xY = \left( \vto\over\voo \right) \xy\,, \qquad
\xW = \xw\,, \qquad
\xZ = \xz\,, \qquad
\xP = \xQ = 0\,, \qquad
\ee
which yields (\ref{Mclassical})
but with $\xx$, $\xy$, $\xz$, and $\xw$ satisfying (\ref{qthree}).
The above solution reduces to the classical case III solution
when $\Lambda_{\cN=1} \to 0$.

\subsection{Summary} \label{endsum}
The various branches of the $\Sp(2N_1){\times}\Sp(2)$ moduli space relevant
to the end of the cascade,
and their type IIB brane interpretations have appeared throughout this section.
Here we collect these results.

The $2{\times}2$ matrices $N_{ij}$ (\ref{Nparam}) mutually commute and their
eigenvalues can be interpreted as the position of the D3-brane probe (and its mirror).

We first summarize the structure of the classical moduli space.
In case I we found $\det N_{ij}=0$ (\ref{Nconifold}),
or equivalently $xy-wz=0$ in the parametrization (\ref{Nparam}),
so the D3-brane probe moves on the orientifolded conifold.
For case II we again found $xy-zw=0$ and
in addition $N_{11}+N_{22}=0$ (\ref{ctwo}), or $w-z=0$.
The latter condition implies that the D3-brane is stuck
to O7-plane/D7-brane stack.
In case III we also found $w-z=0$ together with
$\xx\xy - \xw \xz  =  \lam \voo \vtt$ (\ref{newD3loc}).
The geometrical interpretation of these equations is less clear,
but some suggestions were presented in the text.

For the quantum moduli space we found a similar structure with 
the quantum and classical solutions joining smoothly.
In case I we found $\det N_{ij}=\ep$ (\ref{qonedetN}),
or $xy-wz=\ep$, so the D3-brane probe moves on the
deformed orientifolded conifold.
For case II we again
found $xy-zw=\ep$ and in addition $N_{11}+N_{22}=0$ (\ref{qtwo}),
or $w-z=0$, so the D3-brane is stuck to O7-plane/D7-brane stack.
In case III we also found $w-z=0$ together with
$ \xx \xy - \xw \xz =  \lam \voo\vtt  +  \epsilon$
(\ref{qthree}).
As in the classical case, the geometrical interpretation of
these equations is unclear.

Some insight into the various branches of the quantum moduli space
can be gleaned from the M-theory lift of the type IIA brane configuration
corresponding to the $\Sp(2N_1){\times} \Sp(2N_2)$ gauge theory,
to which we turn next.

\setcounter{equation}{0}
\section{Type IIA and M-theory interpretations} \label{IIA}

In the previous sections we have seen that the moduli space of the
$\Sp(2N_1){\times} \Sp(2N_2)$ gauge theory
and its modification by the ADS superpotential has a richer
structure compared to that of its unorientifolded cousin, the
$\SU(N_1){\times}\SU(N_2)$ gauge theory.

It is fruitful to study the structure of the moduli space of the
$\Sp(2N_1){\times} \Sp(2N_2)$ theory
from the viewpoint of the associated type IIA string theory configuration
and its lift to M-theory, where some of the results obtained in
the previous sections can be understood. We will start by briefly
reviewing the type IIA setup to make the presentation more self contained.

\subsection{Type IIA configurations}

The $\SU(N){\times} \SU(N)$ superconformal gauge theory
with chiral multiplets in the
$2 (\Yfund,\overline{\Yfund}) \oplus 2(\overline{\Yfund},\Yfund)$
representations arises in type IIA string theory
as the world-volume field theory
on D4-branes suspended between two NS5-branes
in an elliptic model (i.e., periodic in the $x_6$ direction)
\cite{Witten:1997}.
There are $N$ D4-branes going
along half the $x_6$ circle,
and $N$ D4-branes going along the other half;
the two stacks of D4-branes give rise to the two factors of
the gauge group.
If the  NS5-branes are parallel,
the $\SU(N){\times} \SU(N)$ gauge theory has $\cN=2$ supersymmetry;
the $\cN=2$ vector multiplet includes a chiral multiplet
in the adjoint representation of the gauge group.
If the NS5-branes are rotated 90 degrees with respect to one another,
the $\SU(N){\times} \SU(N)$ gauge theory has only $\cN=1$
supersymmetry
\cite{Uranga:1998, Dasgupta:1998}.
Rotating the NS5-branes
\cite{Barbon:1997}
corresponds field-theoretically
to including (opposite sign) masses (\ref{masses})
for the adjoint chiral multiplets,
which breaks the supersymmetry to $\cN=1$,
and integrating them out.

The introduction of a pair of orientifold 6-planes into this configuration
results in various $\cN=2$ \cite{Uranga:1998a}
and $\cN=1$ \cite{Uranga:1998,Park:1999, Naculich:2001}
world-volume theories on the D4-branes,
particular examples of which are the models
\be
\label{modelsagain}
\begin{array}{lllll}
&\mathrm{(i)}& \Sp(2N){\times} \Sp(2N)\,,
\qquad &\mathrm{with} \qquad
&2(\Yfund,\Yfund)\oplus 4(\Yfund,1) \oplus 4(1,\Yfund) \non \\
&\mathrm{(ii)}&
\Sp(2N) {\times} \SO(2N{+}2)\,,
\qquad & \mathrm{with}\qquad
&2(\Yfund,\Yfund)
\end{array}
\ee
whose IIB realizations were already discussed in sec.~2.
The O6-planes span the 0123789 directions
and are separated in the (compact) 6 direction;
the two NS5-branes are placed between the O6-planes
and are related to each other by the orientifold symmetry.
If the NS5-branes are parallel,
spanning the 0123 and $v=x_4+i x_5$ directions,
the world-volume field theories have $\cN=2$ supersymmetry
and include chiral multiplets in the adjoint representation
of the gauge group.
The NS5-branes may be rotated
(in opposite directions)
toward the $u=x_8 +i x_9$ plane
(so that one of them spans the $v \cos\alpha + u \sin\alpha$ plane
and the other spans the $v \cos\alpha - u \sin\alpha$ plane)
while still respecting the orientifold symmetry
(which takes $x_6 \to -x_6$ and $v\to -v$).
When $\alpha=\pi/4$,
the NS5-branes become orthogonal,
and the world-volume field theory on the D4-branes is given by
(\ref{modelsagain}).
In model (i), both orientifold planes are O$6^-$ planes;
the configuration also contains 8 D6-branes for cancellation
of 6-brane charge.
In model (ii), there is one O$6^+$ and one O$6^-$ plane and no D6-branes.

The form of the $\Z_2$ orientifold action on the conifold
in the type IIB configuration may be determined \cite{Naculich:2001}
from the rotated IIA brane configuration
described above \cite{Uranga:1998, Dasgupta:1998}.
The D3-branes move in the background
$ x y =  (u\cos\alpha +v\sin\alpha) (u\cos\alpha-v \sin\alpha)$.
When $\alpha=\pi/4$,
this is just a conifold $xy= w z $,
where
$w= {1\over\sqrt2} (u+v)$ and
$z= {1\over\sqrt2} (u-v)$.
The orientifold action implies $u \to u$, $v \to -v$
so that $w \leftrightarrow z$,  as discussed in sec.~2.

Generalizations of the orientifolded models
described above may be obtained by  suspending
$2M$ additional D4-branes between the NS5-branes
that only go along one of the two halves of the $x_6$ circle.
The extra D4-branes break the superconformal
invariance and are the type IIA analog of the
fractional D3-branes in the type IIB theory.
For recent discussions of cascading theories
from the type IIA viewpoint, see \cite{Maldacena:2000,Dasgupta:2001}.

\subsection{M-theory configurations}

Next, we turn to the M-theory lifts of these type IIA brane configurations.
First, consider the configuration
corresponding to the superconformal $\Sp(2N){\times} \Sp(2N)$ gauge theory
with two orthogonal NS5-branes
(one spanning the $z$ plane and the other the $w$ plane)
and $2N$ D4-branes wrapping all the
way around the $x_6$ circle.
Because the D4-branes do not end on the NS5-branes,
but pass through, they can move transversely away
(in the directions $z$, $w$, and $x_7$)
from the NS5-branes.
The motions of each of the $N$ D4-branes
(which are correlated with the motion of the $N$ mirror branes)
together with the Wilson loop expectation value around $x_6$,
gives rise to a six-dimensional moduli space,
which is classically a conifold.
Since the 2 NS5-branes and the D4-branes can be physically separated,
each lifts to a separate M5-brane \cite{Dasgupta:1998}.

Next consider the case $2N_1 > 2N_2$,
in which superconformal symmetry is broken.
$2N_2$ of the branes still wrap all the way around the $x_6$ circle,
and can move transversely away from the other branes;
the classical moduli space of these branes is, as before, the conifold.
These branes lift to a ``toroidal'' M5-brane which is wrapped in the $x_6$
and  $x_{10}$ directions.

There are $2N_1-2N_2$ additional D4-branes
that wrap only half-way around the circle.
These D4-branes end on the two NS5-branes
(which have $w=x_7=0$ and $z=x_7=0$ respectively)
and are therefore pinned in the $z$, $w$,  and $x_7$ directions.
The two NS5-branes and the D4-branes connecting them lift to a single M5-brane
\cite{Dasgupta:1999}.
This M5-brane should be similar to the ``MQCD'' brane
that occurs in the (non-elliptic) type IIA model
with O6-planes \cite{Ahn:1998b,Park:1998}
which gives rise to the $\cN=1$ $\Sp(2N_c)$ model;
in the limit where the $x_6$-periodicity becomes large,
they should become identical.

We briefly describe the form of the MQCD brane
in the $\cN=1$ $\Sp(2N_c)$ model obtained in a model with O6-planes,
following refs.~\cite{Hori:1998, Ahn:1998b}.
Begin with a $\cN=2$ $\Sp(2N_c)$ model
with $2N_f>0$ massless fundamentals
which arises from a IIA configuration with parallel NS5-branes
extended in the $v$ direction.
This configuration lifts to an M5-brane
whose embedding is given by the Seiberg-Witten curve \cite{D'Hoker:1997}
\bea
\label{Spcurve}
t_+ + t_- &=& C(v^2) \ = \ v^{2N_c} + \cdots \\[.1in]
\label{background}
t_+ t_- &=&  \Lambda_{\cN=2}^{4N_c+4-2N_f} v^{2N_f -4}
\eea
(A possible $v^{-2}$ term on the right hand side of the first
equation vanishes because of the masslessness of the fundamental fields.)
To obtain the curve for the $\cN=1$ theory, we must relatively rotate
the NS5-branes, as described above.
This is possible only if the curve (\ref{Spcurve}) degenerates to genus zero,
in which case the coefficients of $C(v^2)$ are fixed.
Rotating the NS5-branes through an angle $\alpha=\arctan(\mutil)$
(where $\mutil$ is proportional to the adjoint mass $\mu$)
in the $v-u$ hyperplane,
we obtain a curve whose projection onto the $v$ plane
is still given by (\ref{Spcurve}),
but with asymptotic behavior
\bea
\label{asymp}
&&x_6 \to -\infty, \qquad u \to \mutil v,
\quad\qquad v \to \infty,  \qquad t_+\to v^{2N_c} \non\\[.1in]
&&x_6 \to +\infty, \qquad u \to -\mutil v,
\,\qquad v \to \infty, \qquad t_- \to v^{2N_c}
\eea
The resulting genus zero curve may be parametrized in terms
of either $ w_+ = u +\mutil v $ or $ w_- = u -\mutil v $.
Letting
\be
v = P(w_+), \qquad t_+ = Q(w_+),
\ee
the orientifold symmetry
$t_+ \leftrightarrow t_-$, $ u\to u$, $v \to -v$ implies
\be
-v = P(w_-), \qquad t_- =  Q(w_-).
\ee
The asymptotic conditions (\ref{asymp}) then imply
\bea
P(w_+) &=& {1\over 2\mutil} \left( w_+  - {w_0^2\over w_+} \right) \\[.1in]
w_+ w_- &=& w_0^2
\eea
for some $w_0$.
Equation (\ref{background}) yields
\bea
Q(w_+) &=& {1\over (2\mutil)^{2N_c} }
w_+^{2N_c +4 - 2N_f} \left( w_+^2  - w_0^2 \right)^{N_f-2}
\eea
where
\be
\label{wnought}
w_0 = 2 \mutil \Lambda_{\cN=2}
\ee
up to a complex phase.
Following the argument of ref.~\cite{Hori:1998},
the parameter $w_0$ 
is proportional to 
the eigenvalue
of the meson matrix constructed from the fundamental fields.

\subsection{Moduli space}

We will now establish the connection between
the configuration of two disconnected M5-branes described above
and the moduli space of the
$\Sp(2N_1) \times \Sp(2)$  gauge theory
as described in sec.~5.
The motion of the toroidal M5-brane,
which is the lift of 2 D4-branes that wrap $x_6$,
is described by  the $2 \times 2$ matrices $N_{ij}$.
The MQCD brane configuration is described by $\MIJ$,
or equivalently $V^{IJ}$.

\medskip
\noindent{{\bf Case I}}
\medskip

In case I, $N_{ij}$ and  $\MIJ$ are unrelated,
which reflects the independence of the 2 M5-branes.
Classically, the moduli space of the toroidal M5-brane
is the conifold (\ref{conifold}).
The ADS superpotential modifies the classical geometry to 
the deformed conifold (\ref{deformedconifold}).

The solution for the antisymmetric meson matrix $V^{IJ}$ (\ref{deBoersoln})
involves a single vev,
which by virtue of the relation \cite{deBoer:1998}
$  \Lambda_{\cN=1}^{3N_1 - 1}  = \mu^{N_1+1} \Lambda_{\cN=2}^{2N_1-2} $
becomes
\be
\tilde{Z}=  2^{ 3-N_1 \over 2N_1 - 2 } \mu \Lambda_{\cN=2}
\ee
This is proportional to the parameter $w_0$ (\ref{wnought})
of the MQCD brane.
This is consistent with our interpretation that $\MIJ$
describes the M5-brane that is the lift of
two orthogonal NS5-branes and $2N_1 -2$ D4-branes.

\medskip
\noindent{{\bf Case II}}
\medskip

In the case II solution, $N_{ij}$ and  $\MIJ$ are also
unrelated,
indicating that the two M5-branes are still disconnected.
$\MIJ$ has the same form as in case I, so the MQCD brane
is unaltered.
In addition to satisfying the deformed conifold constraint,
the $N_{ij}$ must also obey
$N_{11}+N_{22}=0$ (\ref{qtwo}).
This may be understood geometrically as follows.

Case II represents a Higgs branch of the gauge theory
in which the scalar vev $Q_2$ is non-zero.
In the type IIA configuration, this branch corresponds to
D4-branes breaking on the D6-branes
that lie in the interval between the two NS-branes
containing the $2$ D4-branes.
Thus, only the D4-branes that wrap around the $x_6$ circle
(those which lift to the toroidal M5-brane)
can break on the D6-branes.
Since the D6-branes are coincident with the O6-plane
(the fundamental fields have no bare mass),
which is located at $v=0$ (i.e., $w=z$),
the D4-branes can only break on them
if they satisfy $w=z$ as well.
This then implies that the toroidal M5-brane
must satisfy the condition $N_{11}+N_{22}=0$.

\medskip
\noindent{{\bf Case III}}
\medskip

Case III represents a Higgs branch of the gauge theory
in which the scalar vev $Q_1$ is non-zero.
This branch corresponds to
D4-branes breaking on the D6-branes
that lie in the interval between the two NS5-branes
containing the $2N_1$ D4-branes.
Since all the D4-branes can now break on the D6-branes,
the configurations of both M5-branes,
described by $\MIJ$ and $N_{ij}$,
are altered by the $Q_1$ vevs.

As in case II, the D4-branes can only break on the D6-branes
if they satisfy $w=z$,
thus the toroidal M5-brane satisfies $N_{11}+N_{22}=0$ (\ref{qthreemat}).
The remaining $2N_1 -2$ D4-branes were already pinned at the D6-brane
locus, so there is no additional constraint on $\MIJ$.

Finally, since the breaking of the D4-branes on the D6-branes
allows the entire configuration of D4-branes to be interconnected,
the M5-branes to which they lift are no longer disconnected;
this is reflected in the fact that $N_{ij}$ and  $\MIJ$
are no longer independent, but are related by eq.~(\ref{related}).

\section{Summary}
In this paper we have presented a description of the moduli space
of the $\cN=1$ cascading $\Sp(2N_1){\times}\Sp(2N_2)$ gauge theory, 
and the interpretation  of its various branches in terms of
both type IIB and type IIA/M-theory brane configurations. 

In section \ref{Mod} we discussed the (classical) F-term equations
appropriate to the generic case, i.e. without restriction to the end
of the cascade. When the scalar components of the $Q_i$'s do not have
vevs, we argued that the D3-branes move on the orientifolded conifold.
When the vevs of the $Q_i$'s are no longer zero we found that there
are subsectors in which the $N_{ij}$'s are no longer mutually
commuting matrices. In these sectors there does not appear to be a
geometric interpretation of the $N_{ij}$'s as (commuting) coordinates.
However, the vevs of these non-commutative $N_{ij}$'s span (at most)
a $4{\times}4$
subspace of the $2N_2{\times}2N_2$ matrices $N_{ij}$, therefore
for $N_2$ large, one intuitively expects them to
be only a $1/N_2$ effect.

In sec.~\ref{End}, which is the main part of the paper, we
presented an extensive study of the various
branches of the moduli space at the end of the cascade. We studied
both the classical and the quantum versions of the moduli space.
The structure of the moduli space and the dual type IIB
interpretations was summarized in sec.~\ref{endsum}.

In section \ref{IIA} we discussed the moduli space from the
viewpoint of type IIA brane configurations and their lift to M-theory.
The solutions of the quantum F-term equations can be interpreted
in terms of the configuration of two M5 branes,
with $N_{ij}$ corresponding to a toroidal M5 brane
that wraps the $x_6$ direction,
and $M_I{}^{J}$ corresponding to an MQCD brane
that is the lift of the NS5-branes and D4-branes connecting them.
The case III solution in which
$N_{ij}$ and $M_I{}^{J}$ are related (\ref{related})
corresponds to one of the Higgs branches of the theory
in which the two M5 branes are connected.

In a companion paper we will discuss the leading $\al'$-corrections
to the supergravity solution for the orientifolded models
discussed in this paper
(analogous to those considered in ref.~\cite{Frolov:2001}
for the supergravity solution of ref.~\cite{Klebanov:2000a})
and the role of these corrections in the dual field theory.

\begingroup\raggedright\endgroup

\end{document}